
\input phyzzx
\overfullrule=0pt

\Pubnum={
USC-95/019\cr
CERN-TH. 95-167\cr
hep-th/9507032}
\date={June 1995}
\titlepage
\title{Hyperelliptic curves for Supersymmetric Yang-Mills}
\bigskip
\author {
Joseph~A.~Minahan\footnote\star
{minahan@physics.usc.edu}}
\address{Department of Physics,  \break
University of Southern California, Los Angeles, CA 90089-0484 USA}
\andauthor{
Dennis Nemeschansky\footnote\dagger
{dennisn@nxth04.cern.ch. On leave from Physics Department, University of
Southern California, Los Angeles, CA   , USA.}}
\address{Theory Division, CERN\break
CH-1211, Geneva 23, Switzerland}
\bigskip
\abstract{
In this paper we discuss the hyperelliptic curve for
$N=2$ $SU(3)$ super Yang-Mills with six flavors of hypermultiplets.
We start with a generic genus two surface and construct the curve in
terms of genus two theta functions.  From this one can construct the
curve for $m_i=u=0$.  This curve is explicitly dual under a subgroup
of $Sp(4,Z)$ which is not isomorphic to $Sp(2,Z)$.
We then proceed to construct the curve for the general
$SU(3)$ theory and discuss the duality properties of the theory.
The results given here differ from those given previously.
}
\vfill
\endpage

\def\NP{{\it Nucl. Phys.\ }}
\def\PL{{\it Phys. Lett.\ }}

\def\PRL{{\it Phys. Rev. Lett.\ }}

\def\ba{{\bf a}}
\def\bb{{\bf b}}
\def\bz{{\bf z}}
\def\bn{{\bf n}}
\def\vth{\vartheta}
\def\vthab{\vth\left[{\ba}\atop{\bb}\right]}
\def\al{\alpha}
\def\be{\beta}
\def\lam{\lambda}
\def\th{\theta}
\def\Imt{{\rm Im}\tau}
\def\Ret{{\rm Re}\tau}
\def\tu{\tilde u}
\def\tv{\tilde v}

\REF\SWI{N. Seiberg and E. Witten, {\bf hep-th/9407087},
        \NP{\bf 426} (1994) {19}.}
\REF\SW{N. Seiberg and E. Witten, {\bf hep-th/9408099},
        \NP {\bf B431} (1994) {484}.}
\REF\KLTY{A. Klemm, W. Lerche, S. Theisen and S. Yankielowicz,
        {\bf hep-th/9411048}, \PL {\bf B344} (1995) {169}.}
\REF\AF{P.C. Argyres and A.E. Faraggi, {\bf hep-th/9411057},
        \PRL {\bf 73} (1995) {3931}.}
\REF\HO{A. Hanany and Y. Oz, {\bf hep-th/9505075}, {\it On the
	Quantum Moduli Space of Vacua of N=2 Supersymmetric $SU(N_c)$
	Gauge Theories.}}
\REF\APS{P.~C.~Argyres, M.~R.~Plesser and A.~D.~Shapere, {\bf hep-th/9505100},
{\it The Coulomb Phase of N=2 Supersymmetric QCD}.}
\REF\FK{H.~M.~Farkas and I.~Kra, {\it Riemann Surfaces}, Springer-Verlag
(1980), New York.}
\REF\Clemens{C.~H.~Clemens, {\it A Scrapbook of Complex Curve Theory}, Plenum
Press (1980), New York.}

$N=2$ supersymmetric Yang-Mills theories have been shown to have a very
rich structure.  In particular, it has been argued that the exact
effective action can be found from a relatively simple complex curve.
This curve basically contains all of the low energy physics, in that
singularities of the curve describe situations where light particles become
massless.

The original work in this direction discussed $SU(2)$ gauge theories with
$N_f$ hypermultiplets transforming in the fundamental
represention of $SU(2)$[\SWI,\SW].  This work was then extended to first
$SU(n)$ theories with $N_f=0$[\KLTY,\AF], and then most recently to
$SU(n)$ theories with $0<N_f\le2n$[\HO,\APS].
In particular the authors in [\APS]
derive a curve for a general $SU(n)$ theory with $N_f=2n$.  The
curves for $N_f<2n$ can then be found by taking appropriate limits.

However, we believe that the final curve presented in [\APS] is not
correct.  The authors claim that the curve for general $SU(n)$ can be
constructed
out of standard toroidal theta functions.  The basic argument is that
by taking the appropriate limit, one can reduce an $SU(2n)$ theory with $2n$
flavors to an $SU(n)$ theory with $2n$ flavors.  This they argue leads to
constraints on the structure of the $SU(n)$ curve.  Then, to get the final
form of the curve, they take another limit, reducing the $SU(n)$ theory
to an $SU(n-1)$ theory with $2n-2$ flavors.  They then compare
coefficients that are functions of the coupling and then inductively determine
the coefficients for all $SU(n)$ theories starting from $SU(2)$.

While these reductions of  gauge groups
make sense for weak coupling, we see no reason why they should
be exact for strong coupling.   In fact, one can see that there are
troubles just with the reduction of $SU(4)$ with $N_f=4$
to $SU(2)$ with $N_f=4$.  The resulting form of the $SU(2)$ curve in [\APS]
is actually inconsistent with the result of Seiberg and Witten[\SW].
$SU(2)$ has a discrete global parity symmetry that should be present
even in strong coupling.
One can compute the discriminant of the curve in [\SW] and find that it
is invariant under the  parity transformation.  But this is not true
for the curve in [\APS], nor can the symmetry be restored by a redefinition of
the curve variables $y$ and $x$.  This then casts some doubt on the whole
reduction scheme.

In this paper we argue how to construct the general curve for $SU(n)$ with
$n>2$, and we explicitly construct it for $SU(3)$.  We show that
for the $SU(3)$ case, the class of invariant theories is generated
by a subgroup of $Sp(4,Z)$ that is {\it not} $Sp(2,Z)$.  We also show
that this subgroup generates a different fundamental region than the
usual $Sp(2,Z)$ region.  We find that the curve depends on genus
two theta functions, but the period matrix, which is the argument of
the theta functions, is restricted to a particular form.  This will
restrict the number of independent theta functions.  The construction
begins with the special case of $u=\langle\tr(\phi^2)\rangle=m_i=0$.
In this case,
there is only one scale in the theory, so the classical coupling is
the full quantum coupling.  The curve is found by explicitly constructing
a genus two curve with known periods.  Then we allow for nonzero $u$ and
$m_i$, which breaks the conformal invariance of the theory.  Making
some mild assumptions about the possible terms that can appear in the
curve and comparing to the semi-classical duality symmetry of $SU(3)$
allows one to find the final
SU(3) curve.  The result found here differs from that in [\APS].

We begin with a basic discussion of curves for hyperelliptic surfaces with
genus $g\ge2$({\it c.f.} [\FK]).
While there is still no proof that the curve for $N=2$ $SU(n)$ has to be
hyperelliptic, it has so far been succesful at reproducing
known behavior at weak coupling.

The advantage of hyperelliptic surfaces is their great simplification.
The generic hyperelliptic curve for a genus $g=n-1$ surface $\cal M$ is of the
form
$$y^2=\prod_i^{2n} (x-x_i),$$
where the $x_i$ are the branch points (Weierstrass points) of the surface.
There is also associated with this curve a Jacobian variety $J$ which is
the $n-1$ dimensional complex plane modded out by the lattice generated
by the pair $(I,\Omega)$.  $I$ is the $n-1$ dimensional identity matrix and
$\Omega$ is the period matrix.  These lattice elements are found from the
curve by integrating a canonical set of one forms around the noncontractible
loops of the curves.  The usual basis is $\lambda_i=x^{i-1}dx/y$

If the surface is hyperelliptic, then there is a
simple map of the curve into $J$.  Namely, the integrals over the one forms
between branch points are given by the halfway points of the lattice, or
the half-periods.  This then suggests that there should be a map between
the half-periods of $J$ and the branch points.

Suppose that the branch points are ordered $(P_1, P_2 ... P_{2n})$ and let
$\Phi$ be a map $\Phi: {\cal M}\to J$.
We can choose $\Phi(P_1)=0$ and then
define $\Phi(P)$ for any point $P$ on $\cal M$
to be the integral over the normalized canonical set of one forms from
$P_1$ to $P$.
Then per our previous arguments, $\Phi(P_i)$ is a half period in $J$.  Which
half periods is determined by the homology basis one chooses.
We choose the basis such that the integral of the canonical set
of one forms around the branch points $P_{2i-1}$ and $P_{2i}$ gives
$\Omega^{(i)}$ where $\Omega^{(i)}$ is the $i^{th}$ column in $\Omega$ and
the integrals around the points $P_{2i}$ and $P_{2n-1}$ give $e^{(i)}$, where
$e^{(i)}$ is the $i^{th}$ column in $I$.  Note that this particular
choice of integration paths $b_i$ and $a_i$
have the canonical intersection matrices
$a_i\cdot a_j=b_i\cdot b_j=0$, $a_i\cdot b_j=\delta_{ij}$  Therefore, we find
$$\int_{P_{2i-1}}^{P_{2i}} d\lambda = {1\over2}\Omega^{(i)} \qquad\qquad
1\le i<n$$
and
$$\int_{P_{2i}}^{P_{2i+1}} d\lambda = {1\over2}(e^{(i)}-e^{(i+1)})
\qquad\qquad 1\le i<n$$
where the form of this last integral is determined by the intersection
matrices.  We can now determine $\Phi$ at the branch points.  It satisfies
$$\Phi(P_{2i-1})={1\over2}(e^{(1)}-e^{(i)}+\sum_j^{i-1} \Omega^{(j)})
\qquad 1\le i<n$$
$$\Phi(P_{2i})={1\over2}(e^{(1)}-e^{(i)}+\sum_j^{i} \Omega^{(j)})
\qquad 1\le i<n$$
and
$$\Phi(P_{2n-1})={1\over2}(e^{(1)}+\sum_j^{n-1}\Omega^{(j)}), \qquad\qquad
\Phi(P_{2n})={1\over2} e^{(1)}$$

Next consider the $\vartheta$-functions, which are maps of $J$ into
the complex numbers.  The $\vartheta$-functions with characters are
defined as
$$\vartheta\left[{\ba}\atop{\bb}\right]({\bz},\Omega)=
\sum_{{\bn}\in Z^n}
exp(\pi i (\bn+\ba)^T\Omega(\bn+\ba)+2\pi i(\bn+\ba)\cdot(\bz+\bb))$$
The $\vth$-functions have nice transformation properties under the modular
transformations
$Sp(2n-2,Z)$.  The $Sp(2n-2,Z)$ transformations are generated by the set
$$\left[{{A^T}\qquad{0}}\atop{{0}\qquad{A^{-1}}}\right]\qquad\qquad
\left[{1\qquad B}\atop{0\qquad 1}\right]\qquad\qquad
\left[{0\qquad 1}\atop{-1\qquad0}\right],$$
where $A$ is an element of $SL(n-1,Z)$ and $B$ is a symmetric matrix with
integer entries.  Then for a generic matrix of the form
$\left[{A\ B}\atop{C\ D}\right]$, $\Omega$ transforms as
$$\Omega\to(A\Omega+B)(C\Omega+D)^{-1}.$$
In particular, note that for
the transformation of the form $\left[{A^T\ 0}\atop{0\ A^{-1}}\right]$,
$\Omega\to A^T\Omega A$.  Hence we have that
$$\vthab(\bz,A^T\Omega A)=\vth\left[{A\ba}\atop{(A^T)^{-1}\bb}\right]
((A^T)^{-1}\bz,\Omega)\eqn\thtrans$$

We can use the theta functions to find a map of the half periods into the
branch points.  First note that $\vth\left[{\bf0}\atop{\bf0}\right]
(\Phi(P_{2i+1}),\Omega)=0$ for $1\le i<n$.
In fact the $\Phi(P_{2i+1})$ generate the vector of Riemann constants for the
$\vth$-functions.   If we add $\Phi(P_3)$ to $\Phi(P)$ in the argument of
the $\vth$-function, then we find that the $\vth$-function vanishes
at $P_1, P_5, P_7... P_{2n-1}$.  Likewise, if we add $\Phi(P_3)+\Phi(P_{2n})$
to the argument, then we find that the $\vth$-function vanishes at
$P_{2n},P_5,P_7...P_{2n-1}$.  Hence consider the function
$$f(P)={\vth^2\left[{1\ 0\ 0 ...}\atop{1\ 1\ 0 ...}\right](\Phi(P),\Omega)\over
\vth^2\left[{1\ 0\ 0 ...}\atop{0\ 1\ 0 ...}\right](\Phi(P),\Omega)}\eqn\ffun$$
We see that double zeros appear in the numerator at $P_1, P_5, P_7,.. P_{2n-1}$
and they appear in the denominator at $P_{2n}, P_5, P_7... P_{2n_1}$.  Hence
$f(P)$ is a meromorphic function with a double zero at $P_1$ and a double pole
at $P_{2n}$.  Hence, the curve describing this surface is given by
$$y^2=x\prod_{k=2}^{2n-1}(x-x_k)$$
where $x_k=x(P_k)$
and the function $f(P)$ is $f=Cx$, with $C$ a constant.  We are also free to
fix the point $x_2$ to any value, so we choose $x_2=1$.  This then determines
$C$.  Hence the values of the branch points are given by
$$x_k={f(P_k)\over C}={
\vth^2\left[{0\ 0\ 0...}\atop{0\ 1\ 0...}\right](\bf 0,\Omega)
\vth^2\left[{1\ 0\ 0...}\atop{1\ 1\ 0...}\right](\Phi(P_k),\Omega)\over
\vth^2\left[{0\ 0\ 0...}\atop{1\ 1\ 0...}\right](\bf 0,\Omega)
\vth^2\left[{1\ 0\ 0...}\atop{0\ 1\ 0...}\right](\Phi(P_k),\Omega)}\eqn\bpeq$$

Actually, we have to be a little careful, since some of these expressions
involve zero divided by zero.  In these cases, one should instead consider
differentials of these functions.  Then there should exist analogs of
Jacobi's triple product identities to express the derivatives of theta
functions in terms of the other theta functions.

But we do not really need such identities.  Instead, one can shift the
$\vth$-functions by different odd periods, such that the new function
still has a double pole at $P_{2n}$ and a double zero at $P_1$, but
still has $x(P_2)=1$,
and is also now well defined at the point in question.
For instance,
in the genus 2 case,
one finds using \bpeq\ that, $x_3$ and $x_4$ are given by
$$x_3={
\vth^2\left[{0\ 0}\atop{0\ 1}\right]
\vth^2\left[{0\ 0}\atop{0\ 0}\right]\over
\vth^2\left[{0\ 0}\atop{1\ 1}\right]
\vth^2\left[{0\ 0}\atop{1\ 0}\right]}\qquad\qquad
x_4={
\vth^2\left[{0\ 0}\atop{0\ 1}\right]
\vth^2\left[{0\ 1}\atop{0\ 0}\right]\over
\vth^2\left[{0\ 0}\atop{1\ 1}\right]
\vth^2\left[{0\ 1}\atop{1\ 0}\right]}$$
However $x_5$ is not well defined in \bpeq.  However if we instead shift
$\vth\left[{00}\atop{00}\right](\Phi(P))$ by $\Phi(P_5)$ and
$\Phi(P_5)+\Phi(P_{2n})$, we obtain a new function that is well defined at
all points except $P_3$.  But it must be the same function as before, assuming
that it is normalized such that $x_2=1$.  Doing this, one finds that
$$x_5={
\vth^2\left[{0\ 1}\atop{0\ 0}\right]
\vth^2\left[{0\ 0}\atop{0\ 0}\right]\over
\vth^2\left[{0\ 1}\atop{1\ 0}\right]
\vth^2\left[{0\ 0}\atop{1\ 0}\right]}$$

If we redefine $x$ and $y$ such that no terms appear in the denominator
of $x_i$, we find the generic quintic equation for a surface of genus 2
$$\eqalign{y^2=x
&\left(x-\vth^2\left[{0\ 0}\atop{1\ 1}\right]\vth^2\left[{0\ 0}
\atop{1\ 0}\right]
\vth^2\left[{0\ 1}\atop{1\ 0}\right]\right)
\left(x-\vth^2\left[{0\ 0}\atop{0\ 1}\right]
\vth^2\left[{0\ 0}\atop{0\ 0}\right]
\vth^2\left[{0\ 1}\atop{1\ 0}\right]\right)\cr
&\left(x-\vth^2\left[{0\ 0}\atop{0\ 1}\right]
\vth^2\left[{0\ 1}\atop{0\ 0}\right]\vth^2\left[{0\ 0}\atop{1\ 0}\right]\right)
\left(x-\vth^2\left[{0\ 1}\atop{0\ 0}\right]
\vth^2\left[{0\ 0}\atop{0\ 0}\right]\vth^2\left[{0\ 0}\atop{1\ 1}\right]\right)
}\eqn\gtwocurve$$
The advantage of writing the curve this way is that the discriminant is
a modular form.  The discriminant $\Delta$ is defined as
$$\Delta= (a_0)^{2n-2}\prod(x_i-x_j)^2$$
where $a_0$ is the leading coefficient in the polynomial in $x$.  Written
this way, $\Delta$ is invariant under the $SL(2,C)$ transformations
$x\to (ax+b)/(cx+d)$, $y\to y/(cx+d)^n$.
Using the class of identities[\Clemens]
$$\vth^2\left[{00}\atop{00}\right]\vth^2\left[{00}\atop{01}\right]=
\vth^2\left[{00}\atop{11}\right]\vth^2\left[{00}\atop{10}\right]+
\vth^2\left[{10}\atop{00}\right]\vth^2\left[{10}\atop{01}\right]\eqn\thid$$
plus those identities that can be generated by modular transformations on
\thid, one can easily show that the discriminant for \gtwocurve\ is
$$\Delta=\prod_{i=even}(\vth_i)^{12}$$
where the product is over the 10 even genus two $\vth$-functions
This construction of the discriminant generalizes to higher genus
hyperelliptic surfaces.
However, one needs to be mindful that the identities similar to
\thid\ on surfaces with $g>2$ are only true for period matrices that
are compatible with hyperelliptics.  For $g>2$, not all Riemann surfaces
are hyperelliptic.  In fact, for these higher genus surfaces, one will
see identities arise when constructing the hyperelliptic curve.
The values one finds for the branch points depends on what divisor one chooses
to begin the construction.  But in the end, the same result should appear,
even though one will find that the branch points depend on different
combinations of $\vth$-functions.  Hence there must be identities between
these different combinations.

We now wish to compare the period matrix of the surface with the matrix
of couplings arising from the $SU(n)$ $N=2$ gauge theory.  For the
classical gauge theory, non-zero expectation values of $s_k$, where
$$s_k=\langle\tr(\phi^k)\rangle$$
generically break the $SU(n)$ gauge theory to $(U(1))^{n-1}$.  A convenient
basis for the $U(1)$ generators are $T_i=(S_i-S_{i+1})/2$, where $S_i$
is the $U(n)$ generator which has $(S_i)_{jk}=\delta_{ij}\delta_{jk}$.
In this case, we have a matrix of couplings $\psi$ that satisfies
$$\psi_{ij}=\tau\delta_{ij}-{\tau\over2}(\delta_{i,j+1}+\delta_{1+1,j})
\eqn\couplings$$
up to a subgroup of $Sp(2n-2,Z)$ transformations, that act on $\psi$ in the
same way that they act on $\Omega$.  $\tau$ is given by
$$\tau={\theta\over\pi}+i{8\pi\over g^2}\eqn\suncoupling$$
where $g$ is the $SU(n)$ coupling.

Quantum effects will change the general form  of $\psi$ in \couplings\
for generic values of $s_k$ and $m_j$, where $m_j$ are the masses
of the $N=2$ hypermultiplets.  However, consider the case where $m_j=0$
and $s_k=0$ for $k<n$.  If the number of hypermultiplets is $N_f=2n$,
then there is only one scale in the theory, namely $s_n$.  This means
that the $\beta$-function is zero and hence the coupling matrix $\psi$
does not change from its classical value.

If $\psi$ maintains its classical value, then it also has some extra
symmetries that are missing for generic quantum values.  In particular,
there are a class of $Sp(2n-2,Z)$ transformations that leave $\psi$
invariant.  These transformations are generated by the matrix $R$
$$R=\left({A^T\qquad0}\atop{0\qquad A^{-1}}\right)\eqn\Rgen$$
where $A$ is a particular matrix that satisfies $A^n=I$.
Then $\psi$ satisfies $\psi=A^T\psi A$.
$R$ is basically a generator for Weyl reflections.  Since the couplings
are not running, and they started out equal, $\psi$ is invariant under
the reflections.  However, once $s_k\ne0$ for $k<n$, or $m_j\ne0$ for some
$j$, then the couplings will differ.  The Weyl reflections then rotate
these couplings into each other.

So now let us suppose that we have a surface where the period matrix
is invariant under the $Sp(2n-2,Z)$ transformations generated by \Rgen.
Let us concentrate on the special case of $n=3$.  It is straightforward
to find the matrix $A$ that leaves $\Omega$ invariant, where $\Omega$
is
$$\Omega =\left({\omega\qquad-{\omega\over2}}\atop{-{\omega\over2}\qquad\omega}
\right).$$
One finds that $A=\left({0\  -1}\atop{1\ -1}\right)$ is the appropriate
matrix.  For this particular $\Omega$, there must be additional identities
among the even $\vth$-functions.  Using \thtrans\ one finds the identities
$$\eqalign{\vth_1=&\vth\left[{00}\atop{10}\right]=
\vth\left[{00}\atop{01}\right]=\vth\left[{00}\atop{11}\right]\cr
\vth_2=&\vth\left[{10}\atop{00}\right]=\vth\left[{01}\atop{00}\right]=
\vth\left[{11}\atop{00}\right]\cr
\vth_3=&\vth\left[{01}\atop{10}\right]=\vth\left[{10}\atop{01}\right]=
-\vth\left[{11}\atop{11}\right]}\eqn\spthid$$
$\vth_0=\vth\left[{00}\atop{00}\right]$ is invariant under the transformation.
With these identities, one can rewrite the genus 2 curve \gtwocurve\  as
$$y^2=x(x-\vth_1^4\vth_2^2)(x-\vth_1^4\vth_3^2)(x-\vth_0^2\vth_1^2\vth_2^2)
(x-\vth_0^2\vth_1^2\vth_3^2)\eqn\gtwonew$$
Actually, using the identities in \thid\ we can reduce this even more.
Setting $\al=\vth_1^4\vth_2^2$, $\be=\vth_1^4\vth_3^2$ and
$Y=\al^2+\be^2-\al\be$, we can rewrite \gtwonew\ as
$$y^2=x(x-\al)(x-\be)(x-Y/\al)(x-Y/\be)\eqn\gtwo$$

The equation in \gtwo\ describes a hyperelliptic surface with a $Z_3$
symmetry.  We now wish to compare this to the curve that one would
expect for the $SU(3)$ theory with six massless hypermultiplets
and all gauge invariant expectation values zero except for $v=s_3$.
The curves constructed are in sextic form, so we want to find a transformation
of the quintic in \gtwo\ into this more usual form.  Generic arguments
give the form of the sextic to be
$$y^2=(x^3-v)^2-f(\tau)x^6\eqn\sextic$$
where $\tau$ is given in \suncoupling.  The curve is written in this form
in order to compare with previous results, but we will find that it is more
useful to express the curve in a slightly modified form.
To proceed, we set $\Omega$, the period matrix for \gtwo\ equal to $\psi$,
the matrix of couplings.  Hence, there should be an $SL(2,C)$ transformation
such that \gtwo\ is transformed into an equation of the form
$$y^2=a_0x^6+a_3x^3+a_6$$

Under the transformation $x\to (ax+b)/(cx+d)$, $y\to y/(cx+d)^3$, the branch
points are transformed to $x_i\to (dx_i-b)/(a-cx_i)$.
Clearly, we should choose the transformation such that three sets of
branch points are equal up to a cube root of unity, and the other three
points are also equal up to a cube root of unity.  A transformation that
accomplishes this is given by
$$b=-\left({\lam^2\over1-\lam}{Y\over\be-\al}\right)a^{-1}\qquad
c=\left({-\lam^2\over\be+\lam\al}\right)a\qquad
d=\left({1\over1-\lam}{\be+\lam\al\over\be-\al}\right)a^{-1}$$
where $\lam=e^{2\pi i/3}$.  $a$ is chosen such that the equation can
be written as
$$y^2=(rx^3-v)^2-sx^6\eqn\curve$$
where we have now inserted a scale $v$ into the equation. (This will
change the discriminant by a factor of $v^{10}$).  Written in this form,
one finds
$$\eqalign{r=&{\al-\be\over2\al\be}={\vth_1^6\over2}
(2\vth_2^6-3\vth_2^4\vth_3^2-3\vth_2^2\vth_3^4+2\vth_3^6)\cr
&\qquad={\vth_1^6\over2}(\vth_2^2+\vth_3^2)(2\vth_2^2-\vth_3^2)
(2\vth_3^2-\vth_2^2)\cr
s=&-{27\over4}(b-a)^4=-{27\over4}\vth_1^8\vth_2^8\vth_3^8}\eqn\rsvals$$
Using the identity
$$\vth_1^2(\vth_2^2-\vth_3^2)=\vth_2^2\vth_3^2\eqn\thidII$$
which follow from the identities in \thid, $r$ can be further simplified
to
$$r={(\vth_1\vth_2\vth_3)^2\over2}(\vth_2^2+\vth_3^2)
(\vth_1^2+\vth_3^2)(\vth_1^2-\vth_2^2).\eqn\rval$$
In order to compare with the result in [\APS],
we can rescale $x$ such that the curve is in the form
$$y^2=(x^3-v)^2-f(\tau)x^6,\eqn\sexf$$
where $f(\tau)$ is given by
$$f(\tau)={s\over r^2}=-{27}{\vth_1^4\vth_2^4\vth_3^4\over
(\vth_2^2+\vth_3^2)^2(\vth_1^2+\vth_3)^2(\vth_1^2-\vth_2^2)^2}\eqn\ftau$$
Note that this function is different from the one presented in  [\APS]
and also conflicts with the conjecture in [\HO].

Let us examine the transformation properties of $r$, $s$ and  $f(\tau)$.  First
consider a rotation of the $SU(n)$ $\th$ angle such that $\th\to\th+2\pi$.
Hence $\tau\to\tau+2$.  This should leave the theory invariant.  As far as
the period matrix is concerned, we have
$$\Omega\to \Omega + B=\Omega+\left({\ 2\ \ -1}\atop{-1\ \ \ 2}\right)
\eqn\shift$$
The terms on the diagonal transform all $\vth$-functions into themselves,
but the off-diagonal pieces transform $\vth_1\to\vth_1$, $\vth_2\to i\vth_3$
and $\vth_3\to i\vth_2$.  Clearly $f(\tau)$ is invariant under this.

Next consider the dual transformations $\Omega\to C^T\Omega^{-1}C^{-1}$
where 
$C=\left({0\ -1}\atop{1\ \ 0}\right)$.
This corresponds to a transformation
of the coupling matrix
$$\psi\to\left({\ 1\ \ \ -{1\over2}}\atop{-{1\over2}\ \ \ \ 1}\right)
\left({-4\over3\tau}\right).$$
In other words, the dual theory has coupling $\tau'=-4/(3\tau)$.  This
transformation takes $\vth_3\to(-i\sqrt{3}\tau/2)\vth_3$,
$\vth_1\to(-i\sqrt{3}\tau/2)\vth_2$ and  $\vth_2\to(-i\sqrt{3}\tau/2)\vth_1$.
In fact, we also learn something else.  Since $s$ and
$r^2-s$ are both modular forms of weight 24,
then clearly $r^2$ is also a modular form, which
is also clear from \rval.
By inspecting the explicit form for $r$, one sees that $r$ also picks
up a minus sign under the transformation.

These two sets of transformations generate a subgroup of $Sp(4,Z)$ and it
is this subgroup under which the theory is invariant.  The two generators
are
$$S=\left({0\ \ \ C}\atop{C^T\ \ 0}\right)\qquad\qquad
T=\left({I\qquad B}\atop{0\qquad I}\right)$$
and their action on $\tau$ is $S:\ \tau\to-4/(3\tau)$ and $T:\ \tau\to\tau+2$.
We stress that this subgroup is {\it not} $Sp(2,Z)$.  For instance, this
group has $(ST)^6=-1$, but $(ST)^3\ne-1$.  Note that $ST$ generates the
transformation (up to factors involving $\tau$)
$$\vth_1^2\to-\vth_2^2\to-\vth_3^2\to-\vth_1^2.$$
This group also has an interesting
fundamental region.  It is bounded by $-1<\Ret<1$ and $|\tau|>2/\sqrt{3}$.
The region is an orbifold with three singularities of order $Z_2$, $Z_3$
and $Z_6$.  This is similar to the fundamental region for $SL(2,Z)$, except
that in this case, the $Z_3$ point is at infinity, while in the
$SL(2,Z)$ case the $Z_6$ point is at infinity.

Now let us examine the weak coupling behavior of $f(\tau)$.  In the limit
$\Imt\to\infty$ the $\vth$-functions have the limits
$\vth_1\to1$, $\vth_2\to2e^{\pi i\tau/4}$ and
$\vth_3\to2e^{\pi i\tau/4}$.
Hence, in this limit $f(\tau)$ behaves as $f(\tau)\approx -108e^{\pi i\tau}$.
The argument of the exponent has the form expected from instanton
contributions.
However, the coefficient in front of the exponent differs from the one
presented in [\APS].

Before proceeding with the more generic case of nonzero $u$ let
us consider the classical spectrum for the massless $SU(3)$ theory.
This will give a better understanding as to why $\tau$ has the
transformations stated above, as well as giving us clues to what
the final form of the curve should be.  To this end, let the orthogonal
generators of the cartan subalgebra  be
$${\rm diag}(1/2,-1/2,0)\qquad{\rm and}\qquad
{\rm diag}(1/\sqrt{12},1/\sqrt{12},-1/\sqrt{3}).$$  Hence the $U(1)$ charges of
the hypermultiplets with this basis are given by
$$(\pm 1/2,\pm 1/\sqrt{12})e\qquad{\rm and}\qquad
(0,\pm 1/\sqrt{3})e$$
where $e$ is the $SU(3)$ charge.  If the
components of the adjoint scalar are $\phi$ and $\sigma$, then the masses
of the hypermultiplets are given by
$$M=\pm e\phi/2 \pm e\sigma/\sqrt{12}\qquad {\rm and}\qquad
M=\pm e\sigma/\sqrt{3}.\eqn\emass$$
The charges of the monopoles are given by $(\pm2,0)(2\pi/e)$ and
$(\pm 1,\pm\sqrt{3})(2\pi/e)$.  Their masses are
$$M=\pm (4\pi/e)\phi\qquad {\rm and}\qquad
M=(2\pi/e)(\pm\phi\pm\sqrt{3}\sigma).\eqn\mmass$$

Classically, the charges and masses of the hypermultiplets and monopoles
are mapped into each other under the dual transformation
$e\to (8\pi/e)(\sqrt{3}/2)$ and $\phi\to\sigma$, $\sigma\to\phi$.
It is easy to see that the transformation of $e$ corresponds to the
transformation $\tau\to -4/(3\tau)$.  The gauge invariant quantities $u$
and $v$ are given by
$$\eqalign{u&={1\over4}(\phi^2+\sigma^2)\cr
v&={1\over12\sqrt{3}}\sigma(3\phi^2-\sigma^2)}\eqn\uvtrans$$
The transformation of $\phi$ and $\sigma$ leads to the transformations
$u\to u$ and $v^2\to -v^2+4u^3/27$.

Consider now the curve when $u$ is nonzero, keeping the hypermultiplets
massless.  By general arguments we expect the form of the curve to be
$$y^2=(rx^3-tux-v)^2-sx^6\eqn\nzu$$
where $t$ is a function to be determined and $r$ and $s$ are the
functions in \rsvals.  The theory should be invariant
under $2\pi$ shifts of the $\th$ angle, which are generated by $T$.
Since $r$ and $s$ are clearly invariant under this symmetry, then $t$ should
be as well.  Moreover, the theory should be invariant under all transformations
that are conjugate to $T$, in particular the transformation $STS$.  Up to
an overall factor, this transformation maps $\vth_2$ into itself and
exchanges $\vth_1$ with $\vth_3$.
Since $r$ has modular weight 12 under $STS$, then in order for \nzu\ to
have nice modular properties, $t$ should have weight 4 under this
transformation.  Finally, in order to have the correct weak coupling behavior,
the leading
order behavior of $t$ should be $q^{1/2}$ as $q\to 0$, where
$q=e^{\pi i\tau}$.  There are two functions constructed from the
$\vth$-functions that have these properties,
$$g_1=\vth_0\vth_1\vth_2\vth_3\qquad{\rm and}\qquad
g_2={(\vth_1^2-\vth_2^2)(\vth_2^2+\vth_3^2)(\vth_1^2+\vth_3^2)\over
2\vth_0^2},\eqn\gvals$$
thus, we expect $t$ to be either $g_1$ or $g_2$, or perhaps a linear
combination of the two.

However, there is a significant difference between $g_1$ and $g_2$.
Under $S$, the functions transform as
$$g_1\to (9\tau^4/16)g_1\qquad{\rm and}\qquad g_2\to -(9\tau^4/16)g_2.
\eqn\gtran$$
Since $r$ transforms as
$r\to -(3\tau^2/4)^{6}r$, we see that the duality properties
depend on whether $t$ is
$g_1$ or $g_2$.  If $t=g_2$, then
\nzu\ is clearly dual under $S$,
since the extra factors can be reabsorbed into $x$
with no change in $u$ or $v$.  We are also free to make a complex rotation
and implement the transformation $v^2\to -v^2$ and $u\to -u$ along with
the dual transformation.
However, if $t=g_1$, then the transformation is dual, if in addition we
have the transformation
$$v^2\to -v^2\qquad{\rm and}\qquad u\to u.\eqn\qudu$$
The second case is
much closer to the classical dual transformation,  the only difference
being that it is missing a shift in $v^2$ by $4u^3/27$.  Thus it
appears that $t=g_1$ is the better choice.  This also has the
advantage of avoiding $\vth$-functions in the denominator's of the
coefficients.
Given this selection, one then learns that the true quantum duality
transformations are given by \qudu.

We can also show that the classical duality cannot be the true quantum
duality.  The discriminant in \nzu\ is given by
$$\Delta=64(r^2-s)v^6[729(r^2-s)v^4-216rt^3u^3v^2+16t^6u^6]\eqn\disc$$
If $u$ and $v$ have the classical dual transformations, then there is
no choice of $t$ that leaves the form of the discriminant invariant.
It is not even possible to find a $t$ that leaves the piece inside the
square brackets invariant.

Up to now, we have been expressing the curve in terms of $\vth$-functions,
but all coefficients in \nzu\ can be written as polynomials of
the two functions $g_1$ and $g_2$.  Given the form of $t$, this
is perhaps more convenient, with the curve now given by
$$y^2=(g_1^2g_2x^3-g_1ux-v)^2-g_1^4(g_2^2-g_1^2)x^6.\eqn\nzug$$
In fact one can simplify \nzug\ even more by absorbing a factor of
$g_1$ into $x$, leaving the curve
$$y^2=(\xi x^3 -ux-v)^2-(\xi^2-1)x^6,\eqn\nxuxi$$
where $\xi=g_2/g_1$.

The situation becomes more complex when the bare masses of the
hypermultiplets are no longer zero.  The
curve should be chosen such that at weak coupling, the discriminant
is zero if
$$m_i^3-um_i-v=0.$$
We also expect that the curve can be written as a polynomial in the
functions $g_1$ and $g_2$ (or $\xi$)
 and that there exists a one-form $\lambda$ which
has residues that are proportional to $m_i$[\SW].  A curve satisfying
these properties is given by
$$y^2=(\xi x^3+r'x^2\sum_im_i-\tilde u x-\tilde v)^2-
(\xi^2-1)\prod(x+m_i),\eqn\meq$$
where $r'$ is to be determined and where $\tilde u$ and $\tilde v$
are shifted values of $u$ and $v$, where the shifts depend on $m_i$.

It was argued in [\APS] that the one-form $\lambda$ is given by
$$\eqalign{\lambda&={ax\over (\xi^2-1)\prod(x+m_i))}(Zdy-dZ)\cr
&={ax\over y}(dZ-(Z/2)\sum_i{1\over x+m_i})}
\eqn\oneform$$
where
$$Z=\xi x^3+r'x^2\sum_im_i-\tilde u-\tilde v,$$
and $a$ is a constant which is adjusted so that the residues have the
proper values.
Clearly, $\lambda$ has poles at $x=-m_i$, with the
residues proportional to $m_i$.  There is also a pole at $x=\infty$.
The sum of all the residues, including the one at infinity, must be zero
in order that the integral of $\lambda$ is zero
along any contractible loop over the genus two surface.
It is easy to see that this puts no restrictions on $\tilde u$
and $\tilde v$, but sets a condition on $r'$, which one can show satisfies
$$r'={1\over 2}(\xi-1).\eqn\rpval$$
Hence the massive curve is given by
$$y^2=\left(\xi x^3+{1\over2}(\xi-1)x^2\sum_i m_i-\tu x-\tv\right)^2
-(\xi^2-1)\prod_i(x+m_i).\eqn\masscurve$$

Unlike the $SU(2)$ case, the $SU(3)$ curve has no parity symmetry to
help fix the final form of $\tu$ and $\tv$, although even for $SU(2)$ there
is still some ambiguity in these shifts.  Since there does not appear
to be any symmetry gained by shifting $\tu$ and $\tv$, one might consider
setting them to
$\tu=u$ and $\tv=v$.  Another possibility is to choose $\tu$ and $\tv$ such
that the curve reduces to
$$y^2=\left(\xi x^3-\tu x-\tv\right)^2
-(\xi^2-1)\prod_i(x-{1\over6}(1-\xi^{-1})\sum_i m_i+m_i)\eqn\masscurveII$$
after a shift in $x$.
This second choice appears better in that it has
the simplest $S$-duality symmetry.

Under $S$, $\xi$ transforms as $\xi\to-\xi$, $v\to v$ and $u\to -u$ (we have
also included a complex rotation) and a factor of $-1$ is absorbed
into $x$.  Clearly it is necessary to transform $m_i$ as well in order
to keep the form of the curve invariant.  A simple calculation shows that
the curve in \masscurveII\ is invariant if $m_i$ transforms as
$m_i\to -m_i+{1\over3}\sum m_i$.
Hence the duality symmetry seems to be much simpler than the $SU(2)$ case,
where there was a complicated triality symmetry that goes along with
the dual transformations[\SW].

While the simple dual structure in \masscurveII\
is not necessarily a proof that this
is the correct $SU(3)$ curve, we can at least show that it is consistent
with other behavior.
If we let $m_6\to\infty$, keeping $-108e^{\pi i\tau}m_6=\Lambda_5$ fixed,
then the theory reduces to $SU(3)$ with five flavors.  The curve, after
rescaling is
$$y^2=(x^3-ux-v)^2-\Lambda_5\prod_i(x-\Lambda_5/12+m_i).\eqn\fivecurve$$
We can then let $m_5\to\infty$ such that $\Lambda_4^2=\Lambda_5m_5$ is
kept fixed.  The curve is now
$$y^2=(x^3-ux-v)^2-\Lambda_4^2\prod_i(x+m_i).\eqn\fourcurve$$
Notice that there is no $\Lambda_4$ dependence inside the parentheses, even
though shifting $u$ by $\Lambda_4^2$ violates no symmetries.  This
result is sensible when one considers the reduction of $SU(3)$ to
$SU(2)$.  Classically, $SU(3)$ is reduced to $SU(2)$ by letting $u$, $v$
and $m_i$ become large, scaling them such that $u=3a^2+U$, $v=-2a^3+2aU$,
and $m_i=-a+M_i$ as $a\to\infty$, with $U$ the $SU(2)$ quadratic casimir
and $M_i$ the $SU(2)$ masses.   We expect this behavior to persist
in weak coupling.

Under an $SL(2,C)$ transformation, one can map the cubic $N_f=4$ curve
in [\SW] to the quartic
$$y^2=\left(x^2(\th_1^4+\th_3^4)+x\th_3^4\sum_iM_i-U+
{1\over2}\th_1^4\sum_{i<j}M_iM_j\right)^2
-4\th_1^4\th_3^4\prod_i(x+M_i),\eqn\sutwofour$$
where $\th_1$ and $\th_3$ are standard $Sp(2,Z)$ theta functions.
The curve in \fourcurve\ reduces to \sutwofour\ if $u=3a^2+s$, $v=-2a^3+2as$,
$M_i=-a+\mu_i$ and $\Lambda_4^2=la^2$, where
$$s={U\over \th_1^4+\th_3^4}-
{\th_1^4/2\over \th_1^4+\th_3^4}\left({\th_1^4/2\over \th_1^4+\th_3^4}
\left(\sum M_i\right)^2
+\sum_{i<j}M_iM_j\right),\eqn\stwoeq$$
$$\mu_i=M_i-{\th_1^4/2\over \th_1^4+\th_3^4}\sum M_i,\eqn\mueq$$
$$l={36\th_1^4\th_3^4\over \th_1^4+\th_3^4},\eqn\leq$$
Now if
\fourcurve\ had  a shift in $u$ of a constant multiplied by $\Lambda_4^2$,
we would have had to subtract this term right back off again in order to
reduce it to the $SU(2)$ curve.  Since $\Lambda_4^2$ is of order $a^2$,
this term will be of the same order as $u$ and hence would imply
that there is really no regime where the classical reduction of $SU(3)$
to $SU(2)$ can take place.
This seems unlikely\footnote\dagger{The authors in [\HO] argue that such
shifts should appear in order to have a singularity structure with
multiplicities $(n,n,N_f-n)$ when $s_i=0$, saying that there is
a possible global symmetry $SU(n)\times SU(n)\times SU(N_f-n)$.  But
it is not clear where such a symmetry could come from.}.  In fact
we can make this argument for any $SU(n)$ curve with $N_f=2n-2$, that is,
the scale $\Lambda_{2n-2}$ only appears in front of the product
$\prod(x+m_i)$.  The one exception is $SU(2)$, where the $N_f=2$ curve is
$$y^2=(x^2-u+\Lambda^2/8)^2-\Lambda^2(x+m_1)(x+m_2).\eqn\sutwotwo$$
This shift in $u$ is necessary to preserve parity.  But of course there
is no other nonabelian gauge group that $SU(2)$ can reduce to, so this is not
a contradiction to the previous argument.

Notice that $s$ and $\mu_i$ reduce to $U$ and $M_i$ in the weak coupling
limit, but that there is some contribution of the masses to $s$, and that
these contributions show up in strong coupling.  This is a
scenario that was  not actually allowed for in [\APS], but must actually occur
in order to reduce $SU(3)$ to $SU(2)$.  In fact the curve in [\APS] is
not consistent with parity.

This constraint on the $N_f=4$ curve still does not uniquely pick the
shifts in $u$ and $v$, although it does rule out some possible shifts.
For example, a shift of the form $\tu=u+A(\xi-1)\prod_{i<j}m_im_j$
is not allowed, since the shifted term would survive down to $N_f=4$.
On the other hand, a term of the form $A(\xi-1)^2\sum_i m_i^2$
survives down to $N_f=5$ but not $N_f=4$.
Of course these shifts do not actually alter the monodromies, they just
shift the singularities to different values of $u$ and $v$.

The generalization to higher $SU(n)$ is possible, but certainly much
messier.  It is still true that the number of independent theta functions
is less than the number of even theta functions.  There are also a host
of identities that allow one to greatly simplify the curve.
We hope to address this and other issues in a subsequent paper.



\ack{J.A.M. thanks Nick Warner for many helpful discussions.  D.N
would like to thank S. Yankielowicz for discussions.  This research
was supported in part by D.O.E.~grant DE-FG03-84ER-40168.}
\vfill
\eject
\refout

\end